
\input harvmac
\def\inbar{\,\vrule height1.5ex width.4pt depth0pt}
\font\cmss=cmss10 \font\cmsss=cmss10 at 7pt
\def\C{\relax\hbox{$\inbar\kern-.3em{\rm C}$}}
\def\Z{\relax\ifmmode\mathchoice
{\hbox{\cmss Z\kern-.4em Z}}{\hbox{\cmss Z\kern-.4em Z}}
{\lower.9pt\hbox{\cmsss Z\kern-.4em Z}}
{\lower1.2pt\hbox{\cmsss Z\kern-.4em Z}}\else{\cmss Z\kern-.4em Z}\fi}
\def\Q{\relax\hbox{$\inbar\kern-.3em{\rm Q}$}}
\def\R{\relax{\rm I\kern-.18em R}}

\def\GL{{\rm GL}}

\def\Diff{\hbox{\rm Diff}}
\def\Aut{\hbox{\rm Aut}}
\def\qbin#1#2{{\left[{#1\atop #2}\right]_q}}
\def\uu{{U_q(sl_2(\C))}}
\def\uq{{U_q^{red}(sl_2(\C))}}
\def\uqk{{U_{q,K}^{red}(sl_2(\C))}}
\Title{}
{\vbox{\centerline{Topological Representations of $\uu$ on the Torus}
\vskip2pt\centerline{and the Mapping Class Group}}}

\centerline{M. Crivelli${}^1$, G. Felder${}^1$, C. Wieczerkowski${}^2$}
\bigskip\centerline{${}^1$Mathematik, ETH-Zentrum, 8092 Z\"urich, Switzerland}
        \centerline{${}^2$Theoretische Physik 1, Universit\"at M\"unster,}
        \centerline{4400 M\"unster, Germany}
\vskip 1cm
\baselineskip 14pt
\noindent
{\it Abstract:} We compute the mapping class
group action on cycles on the configuration
space of
the torus with one puncture, with coefficients in a local
system arising in conformal field theory.  This action
commutes with the topological action of the quantum
group $\uu$, and is given in vertex form.
\baselineskip=14pt
\newsec{Introduction}
\def\np{Nucl. Phys. }

%
\nref\REFFW{G. Felder
and C. Wieczerkowski, Topological representations of $U_q(sl_2)$,
Commun. Math. Phys. 138, 583-605 (1991)}
\nref\REFSV{V.V. Schechtman and A.N. Varchenko, Quantum groups and homology
of local systems, ICM-90 Satellite Conference Proceedings,
Algebraic Geometry and Analytic Geometry, Springer (1991)\semi
Arrangements of hyperplanes and Lie algebra homology,
Inv. Math. 106, 139-194 (1991)}
\nref\REFLawr{R.Lawrence, A topological approach to representations
of the Iwahori-Hecke algebra, Int. J. Mod. Phys.
A5, 3212-3219 (1990)}
\nref\REFCFW{M. Crivelli, G. Felder and C. Wieczerkowski, Generalized
	hypergeometric functions on the torus and the adjoint action
	of $U_q(sl_2)$, Commun. Math. Phys. 154, 1-23 (1993)}
\nref\REFDrin{ V.G. Drinfeld,
``Quantum groups,'' Proceedings of the International
Congress of Mathematicians, p.~798-820, Academic Press (1986).}
\nref\REFLusz{G. Lusztig, Finite dimensional Hopf algebras arising from
quantum groups, J. Amer. Mat. Soc. 3, 257-296 (1990)}
\nref\REFFK{J.Fr\"ohlich, T.Kerler, Quantum groups, quantum categories,
and quantum field theory, Springer Lecture Notes (to appear)}
\nref\REFRT{N.Reshetikhin, V.Turaev, Invariants of 3-manifolds via link
polynomials and quantum groups, Inv. Math. 103, 547-597 (1991)}
\nref\REFBPZ{A. A. Belavin, A. M. Polyakov and A. B.
Zamolodchikov, Infinite conformal symmetry in two-dimensional
conformal field theory, Nucl. Phys. B241, 33-380 (1984)}
\nref\REFDF{ V.S. Dotsenko and V.A. Fateev,
Conformal algebra and multipoint
correlation functions in 2d statistical models,
\np B240[FS12], 312-348 (1984)}
\nref\REFFZ{ A.B. Zamolodchikov and V.A. Fateev, Operator
algebra and correlation
functions in the two-dimensional $SU(2)\times SU(2)$ chiral Wess-Zumino
model, Sov. J. Nucl. Phys., 43, 657-664 (1986)}
\nref\REFFelNP{G. Felder, BRST approach to minimal models, Nucl. Phys. B317,
215-236 (1989)}
\nref\REFBernardFelder{D. Bernard and G. Felder,
Fock representations and BRST cohomology in SL(2) current algebra,
Commun. Math. Phys. 127, 145-168 (1990)}
\nref\REFSVI{V. V. Schechtman and A. N.
Varchenko, Integral representations of N-point conformal correlators
in the WZW model, MPI/89-51 (1989)}
\nref\REFGAW{K. Gaw\c edzki, Quadrature of conformal field theory,
Nucl. Phys. B328, 733 (1989)}
\nref\REFGK{K. Gaw\c edzki and A. Kupiainen, SU(2)
Chern-Simons theory at genus zero, Commun. Math. Phys. 135, 531-546
(1991)}
%
%
\nref\REFFelSil{G. Felder and R. Silvotti, Modular covariance of
	minimal model correlation functions, Commun. Math. Phys. 123,
	1-15 (1989)}
\nref\REFBirman{J.S. Birman, Braids, Links, and Mapping Class Groups,
	Annals of Mathematics Studies 82, Princeton University Press (1974)}
%
%
%
%
%
\nref\REFMajid{V. Lyubashenko and S. Majid,
 Braided groups and quantum Fourier transforms,
preprint DAMTP/91-26}
%

We consider topological representations of $\uu$
appearing in free field representations of conformal field theories on
the torus based on $SU(2)$.
Topological representations of quantum  groups
on the complex plane were introduced in
[1,2,3].
The torus has been investigated in [4].

Fock space traces of products of vertex operators yield
multivalued holomorphic differential forms on configuration spaces
over the torus.
Quantum groups [5-8] enter through their action on
a certain space $A$ of linear forms
on the space of holomorphic
multivalued differential forms with  given monodromy.
These forms are given by integration on products of
loops. Singular vectors with respect with this action
give cycles, and define thus linear forms on cohomology.
We consider the torus with one puncture together with the local
system given in [4], associated to the monodromy of the differential
forms.
We restrict our attention to the quantum group $\uu$ at $q$
a $2p$'th root of unity.
The topological action of $\uu$ has been identified in [4]
with the adjoint representation in the sense that the
space $A$ is
isomorphic to $\uu$ as $\uu$-module with the adjoint action.
The main input from conformal field theory [9-17] is the
form of the local system.

An important feature of conformal field theories on the torus is
modular invariance [17].
A natural question to pose is the meaning of
modular transformations on the side of topological representations.
The first observation is that the local system
coming from conformal field theory is compatible with modular
transformations in a sense to be defined below.
As a consequence the modular group acts on the space $A$.
The second observation is that we can explicitly compute
the action of the modular group on $A$
by contour deformation methods.
Using the identification of $A$
with the quantum group algebra, we obtain the action of the
modular group on the latter.

Since the action of the modular group commutes with the
topological action of $\uu$, it commutes on the algebraic side
with the adjoint action.
The result is a ``vertex'' form of the
generators $T$ and $S$ of the modular group. These
generators are expressed in terms of the $R$-matrix
of an enlarged version of $\uu$ (the ``$K$-generated
algebra'') and the Haar measure on $\uu$.
A representation of the mapping
class group in ``SOS'' form
arises in the study of three-manifold
invariants of Reshetikhin and Turaev [8].

We obtain as a byproduct
the quantum group interpretation of modular invariance.
Namely, the action of the modular group leaves invariant
the subspace of singular vectors in the adjoint representation.

It turned out that very similar formulas have been discovered
independently,
in the context of braided groups of monoidal categories,
by Lyubashenko and Majid [18].
\newsec{Configuration spaces and local systems on the torus}
Let $X$ be a torus with one puncture.
\subsec{Representation of $X$}
Let $D_{R}(w)=\{ z\in\C | |z-w|<r\}$, the open disc. We represent
$X=(\overline{D_{R}(0)}\setminus \cup_{i=1}^{2}D_{R'}(w_{i}))/\sim$,
the disc with two holes, the boundaries of which we identify. E.g., we
take $w_{1}=-{R\over 3}$ and $w_{2}={R\over 3}$, define
$\phi (w_{1}+R' e^{i(\varphi -\pi )})=w_{2}+R' e^{-i\varphi}$, and
identify $\phi (z)\sim z$. Let $p_{0}=-R$ serve as a base point.
$\pi_{1}(X,p_{0})$
is generated by elements $\alpha$ and $\beta$ as represented by the
loops in $X$ based at $p_{0}$ shown in figure (1). For later purpose,
we introduce the abbreviations $\alpha'=\beta^{-1}\circ\alpha^{-1}\circ\beta$
and $\gamma=\alpha\circ\beta^{-1}\circ\alpha^{-1}\circ\beta$.
\subsec{Configuration spaces and braid groups on $X$}
For $r\geq 1$, we define configuration spaces
\eqn\Ii
{X_{r}=\left( X^{r}\setminus\cup_{1\leq i<j\leq r}\{
(z_{1},\dots,z_{r})\in X^{r}|z_{i}=z_{j}\}\right)/S_{r}.}
$S_{r}$ is the symmetric group, acting from the right. Let
$*_{r}=[x_{1},\dots,x_{r}]$ be a base point in $X_{r}$.
$\pi_{1}(X_{r},*_{r})$ is the braid group with $r$ strings on $X$.
We call a base point admissible if $-R\leq x_{1}<\dots <x_{r}\leq
-{R\over 3}-R'$.
Braid groups defined with respect to different
admissible base points are canonically isomorphic. We will always
assume $*_{r}$ to be an admissible base point such that
$\{ x_{1},\dots, x_{r}\}\subset\partial X$.
$\pi_{1}(X_{r},*_{r})$ is
generated by elements $\sigma_{i}$, $1\leq i\leq r-1$, $\alpha$, and
$\beta$.
Intuitively, $\sigma_{i}$ interchanges $x_{i}$ with $x_{i+1}$
counterclockwise, while $\alpha$ and $\beta$ move $x_{r}$ along the
respective loops, other components of $*_{r}$ being kept fixed.
An abundance of relations hold among these generators. We will not
present them here.
\subsec{Local systems over $X_{r}$}
Let $p$ be an odd positive integer. Put $q=e^{{{\pi i}\over p}}$ and
define $2p$ by $2p$ matrices $A$ and $B$ with entries
\eqn\Iii{\eqalign{
A_{m,n}&=q^{1-m}\delta_{m,n},\cr
B_{m,n}&=\sum_{l\in\Z}\delta_{m,n+2pl+1}.\cr}}
They satisfy $AB=q^{-1}BA$. Let $V=\C^{2p}$. The assignments
$\rho_{r}(\sigma_{i})=-q^{2}$,
$1\leq i\leq r-1$, $\rho_{r}(\alpha)=A^{2}$, and
$\rho_{r}(\beta)=B^{2}$ define a $2p$-dimensional representation
$\rho_{r}:\pi_{1}(X_{r},*_{r})\rightarrow\GL(V)$.
It is the monodromy
representation associated with multivalued differential
forms on $X_{r}$ mentioned above.
$\rho_{r}$ is the direct sum of two equivalent $p$-dimensional irreducible
representations.

Let $X_{r}^{0}$ be the subspace of $X_{r}$ consisting of
configurations which contain $p_{0}$ among their components.
We then define $\phi_{r}:X_{r-1}\setminus X_{r-1}^{0}\rightarrow
X_{r}^{0}$
to be the bijection which inserts $p_{0}$. The family of
representations
$\rho_{r}$, $r\geq 1$, is compatible in the following sense.
Let $\pi_{1}(\phi_{r}):\pi_{1}(X_{r-1}\setminus
X_{r-1}^{0},*_{r-1})\rightarrow \pi_{1}(X_{r}^{0},\phi_{r}(*_{r-1}))$
be the isomorphism induced by $\phi_{r}$, then
$\rho_{r}\circ\pi_{1}(\phi_{r})=\rho_{r-1}$.

With $\rho_{r}$ we associate the local system $L_{r}(X)=
\hat{X}_{r}(*_{r})
\otimes_{\pi_{1}(X_{r},*_{r})}V$, a flat vector bundle over $X_{r}$
with distinguished trivialization over $*_{r}$, the holonomy
associated with elements of $\pi_{1}(X_{r},*_{r})$ being $\rho_{r}$.
Due to the compatibility, $\phi_{r}$ can be lifted to
$L_{r}(\phi_{r}): L_{r-1}(X)\big|_{X_{r-1}\setminus
X_{r-1}^{0}}\rightarrow L_{r}(X)\big|_{X_{r}^{0}}$.
We define $L_{r}(\phi_{r})([x,v])=[\phi_{r}(x),v]$,
which is checked to be well defined.
\newsec{Topological representations of $\uq$}
We summarize briefly the constructions leading to topological
representations of $\uq$ adjusting the notations to the
present setup.
\subsec{Families of nonintersecting loops with values in the local
system}
Let $Q_{r}=]0,1[^{r}\cup\bigcup_{i=1}^{r}]0,1[\times\dots\times\{0,1\}
\times\dots\times]0,1[$ and $\gamma_1,\dots,\gamma_r$ be
loops $[0,1]\to X$ starting and ending at $p_0$, nonintersecting
except at $p_0$. Define $[\gamma_1,\dots,\gamma_r]:Q_r\to X_r$ be
the corresponding embedding. Denote by
$[\beta^{j},\alpha^{k}]:Q_{r}
\rightarrow X_{r}$, $j+k=r$  a family of nonintersecting loops
obtained by homotopic deformation of $j$ $\beta$-loops and $k$
$\alpha$-loops given by
\eqn\IIi
{[\beta^{j},\alpha^{k}](t_{0},\dots,t_{r-1})=
[\beta^{(0)}(t_{0}),\dots,\beta^{(j-1)}(t_{j-1}),
\alpha^{(j)}(t_{j}),\dots, \alpha^{(r-1)}(t_{r-1})].}
It represents a locally finite $r$-chain in $X_{r}$ with boundary in
$X_{r}^{0}$.

We lift it to take values in $\hat{X}_{r}(*_{r})$. We specify the lift
by choosing an admissible point on its image, connecting this point to
the base point by an admissible path. Then the equivalence class
$[[\beta^{j},\alpha^{k}],v]$, $v\in V$, defines a family of
nonintersecting loops in $X$ with values in $L_{r}(X)$.
The space of families of nonintersecting loops in $X$ with values in
$L_{r}(X)$ is denoted by $A_{r}(X_{r},X_{r}^{0};L_{r})$
or shorter by $A_{r}$. Its precise
definition contains equivalence relations reflecting the possibility
of homotopic deformation, reparametrization, and splitting of loops
(see [1]).
The elements $[[\beta^{j}, \alpha^{k}],e_{n}]$, $0\leq j,k\leq
\min\{r, p-1\}$ such that $j+k=r$, and $1\leq n\leq 2p$, constitute a
basis. Here $(e_{n})_{m}=\delta_{n,m}$. A family which contains
$p$ homotopic loops is put equivalent to zero. Therefore, we restrict
ourselfs to $r\leq 2p-2$.
\subsec{Topological action of $\uq$}

The basic ingredience of
topological representations are
operators $E:A_{r}\rightarrow A_{r-1}$,
$F:A_{r}\rightarrow A_{r+1}$, and $K^{2}:A_{r}\rightarrow A_{r}$
defined by
\eqn\IIii{\eqalign{
E[[\beta^{j},\alpha^{k}],e_{n}]&=
-L_{r}(\phi_{r})^{-1}\partial[[\beta^{j},\alpha^{k}],e_{n}],\cr
F[[\beta^{j},\alpha^{k}],e_{n}]&=
q^{-2j-2k-2}[[\beta^{j},\alpha^{k},\gamma],e_{n}],\cr
K^{2}[[\beta^{j},\alpha^{k}],e_{n}]&=
q^{-2r-2}[[\beta^{j},\alpha^{k}],e_{n}].\cr}}
They are shown to satisfy the relations of $\uu$:
$K^2E=q^2EK^2$,$K^2F=q^{-2}FK^2$, $[E,F]=K^2-K^{-2}$, and
the additional relations
$E^p=F^p=(K^2)^{2p}-1=0$, defining $\uq$.

Thus $\bigoplus_{r=0}^{2p-2}A_{r}$ comes equipped with the structure of
a module over $\uq$. We identify this representation
as the adjoint representation. Let
$\phi:\bigoplus_{r=0}^{2p-2}A_{r}\rightarrow \uq$ be the
map
\eqn\IIiii{\eqalign{
\phi [[\beta^{j},\alpha^{k}],e_{n}]&=
N(j,k,n) F^{k}T_{n-1}E^{p-1-j},\cr
N(j,k,n)&=(-1)^{j}q^{2j+2n}q^{{1\over 2}j(j-1)}
{[j]_{q}!\over [1]_{q}^{j}}
q^{(j+k)(j+k-1)+(j+k)(1-n)+j(1+n)}.\cr}}
An explicit computation proves that $\phi$ is a homomorphism of
$\uq$ modules. Moreover, it is one-to-one and
onto. Here $T_{n}={1\over 2p}\sum_{m=0}^{2p-1}q^{-nm}K^{2m}$.

The actions of $\uq$ on itself by left multiplication
and by right multiplication, twisted with the antipode,
also have topological counterparts. The operator which implements
left multiplication by $F$ is called $F_{L}$. The operator which
corresponds to right multiplication by $\eta (F)$ is denoted by
$F_{R}$. On the topological side they are given by
\eqn\IIiv{\eqalign{
F_{L}[[\beta^{j},\alpha^{k}],e_{n}]&=
q^{n-1-2j-2k}[[\beta^{j},\alpha^{k+1}],e_{n}],\cr
F_{R}[[\beta^{j},\alpha^{k}],e_{n}]&=
q^{-2j-2k-2}[[\alpha',\beta^{j},\alpha^{k}],e_{n}].\cr}}
Intuitively, $F$ adds a $\gamma$-loop, while $F_{L}$ and $F_{R}$
add $\alpha$- and $\alpha'$-loops respectively. The interpretation
of $\bigoplus_{r=0}^{2p-2}A_{r}$ as a bimodule will not be worked
out here.

The important formula of this section to keep in mind is \IIiii .

\newsec{Action of the Mapping Class Group}

Let $\Diff(X)$ be the group of diffeomorphisms which leave $\partial X
=\{z\in\C\,|\,|z|=R\}$ invariant. Let $\Diff_0(X)$ be the subgroup of
$\Diff(X)$ consisting of diffeomorphisms homotopic to the identity.
The mapping class group of $X$ is defined as
\eqn\IIIi{
{\cal M}_{1,1}(X):=\Diff(X)/\Diff_0(X).
}
A reference on mapping class groups is \REFBirman.

\subsec{Generators of ${\cal M}_{1,1}(X)$}

${\cal M}_{1,1}(X)$ is generated by Dehn twists. On the torus, we have two
kinds of Dehn twists, $T_\alpha$ and $T_\beta$. $T_\alpha$ is defined as
follows. Consider the annulus $\{z\in X\,|\,r+\epsilon\leq|z-w_1|\leq
r+2\epsilon\}$ with $\epsilon>0$ fixed. We define a map of this annulus to
itself by $T_\alpha(w_1+z):=w_1+e^{i\varphi(|z|)}z$ with $\varphi$ a
smooth function interpolating between $\varphi(r+\epsilon)=0$ and
$\varphi(r+2\epsilon)=2\pi$. We say that $T_\alpha$ is the Dehn twist
associated with the loop $t\mapsto w_1-(r+2\epsilon)e^{2\pi it}$, $t\in
[0,1]$.
The Dehn twist $T_\beta$ is associated with the loop $t\mapsto(w_1+r)
(1-t)+(w_2-r)t$. See figure (3). The orientations we use are shown by arrows.
$T_\alpha$ and $T_\beta$ leave the base point $*_r$ invariant. (Recall
that $*_r$ is a configuration on $\partial X$.) Thus we
have a map from ${\cal M}_{1,1}(X)$ to $\Aut(\pi_1(X_r,*_r))$, the
automorphisms of $\pi_1(X_r,*_r)$, $r\geq 1$.

\subsec{Compatibility of local systems}

We define a representation $\rho_r:\pi_1(X_r,*_r)\to \GL(V)$ to be
compatible with ${\cal M}_{1,1}(X)$ if $\rho\circ T\simeq\rho$ for all
$T\in{\cal M}_{1,1}(X)$. This means that for every $T$ there exists
a matrix $D(T)\in \GL(V)$ such that
\eqn\IIIii{
\rho\circ T(\sigma)=D(T)\rho(\sigma)D(T)^{-1}.
}
If a representation $\rho_r$ is compatible with ${\cal M}_{1,1}(X)$,
then we have an action of ${\cal M}_{1,1}(X)$ on $L_r$ given by
\eqn\IIIiii{
L_r(T):L_r\to L_r,\;[x,v]\mapsto[T(x),D(T)v],
}
which is well defined due to equation \IIIii.

The family of local systems $\rho_r:\pi_1(X_r,*_r)\to \GL(V)$, $r\geq 1$,
is indeed compatible with ${\cal M}_{1,1}(X)$.
For the generators $T_\alpha$ and $T_\beta$ the compatibility is
proved by defining
\eqn\IIIiv{
\eqalign{
D(T_\alpha)&:=\sum_{l=0}^{2p-1}q^{{1\over2}l(l-2)}A^l, \cr
D(T_\beta)&:=\sum_{l=0}^{2p-1}q^{-{1\over2}l(l+2)}B^l.
}}
and by noting that the action of $T_\alpha$ has the form
$T_\alpha(\alpha)=\alpha$ and $T_\alpha(\beta)=\alpha\circ\beta$,
while that of $T_\beta$ has the form $T_\beta(\alpha)=\alpha\circ\beta$
and $T_\beta(\beta)=\beta$.
Note also that $D(T_\alpha)$ and $D(T_\beta)$ have matrix elements
\eqn\IIIv{
\eqalign{
D(T_\alpha)_{m,n}&=\left\{\sum_{l=0}^{2p-1}q^{{1\over2}l^2}\right\}
q^{-{1\over2}m^2}\delta_{m,n}, \cr
D(T_\beta)_{m,n}&=\sum_{k\in\Z}\sum_{l=0}^{2p-1}q^{-{1\over2}l(l+2)}
\delta_{m,n+2pk+l}.
}}

\subsec{Action of ${\cal M}_{1,1}$ on $A_r(X_r,X_r^0;L_r)$}

The mapping class group ${\cal M}_{1,1}$ acts on $A_r$ as
follows:  $T\in{\cal M}_{1,1}$ acts by
\eqn\act{
L_r(T)[[\gamma_1,\dots,\gamma_r],v]
=[T\circ[\gamma_1,\dots,\gamma_r],D(T)v].
}
We compute the action of $T_\alpha$ and $T_\beta$ on the basis elements
$[[\beta^j,\alpha^k],e_n]$ of $A_r(X_r,X_r^0;L_r)$.

\bigskip
\noindent{\it Action of $T_\alpha$}
\bigskip

Let us define $\delta:=\alpha\circ\beta$. Let $r=j+k$. The action of
$T_\alpha$ on $[[\beta^j,\alpha^k],e_n]$ is seen to have the form
\eqn\IIIvi{
L_r(T_\alpha)\bigl([[\beta^j,\alpha^k],e_n]\bigr)
=[[\delta^j,\alpha^k],D(T_\alpha)e_n].
}
The first problem is to decompose \IIIvi\ in terms of the basis of
$A_r(X_r,X_r^0;L_r)$. The decomposition is performed with the help of
\eqn\IIIvii{
\eqalign{
&[[\beta^j,\delta^{l+1},\alpha^k],e_n] \cr
&\qquad=[[\beta^{(0)},\ldots,\beta^{(j-1)},\delta^{(j)},\ldots,\delta^{(j+l)},
\alpha^{(j+l+1)},\ldots,\alpha^{(j+l+k)}],e_n] \cr
&\qquad=[[\beta^{(0)},\ldots,\beta^{(j)},\delta^{(j+1)},\ldots,\delta^{(j+l)},
\alpha^{(j+l+1)},\ldots,\alpha^{(j+l+k)}],e_n] \cr
&\qquad\qquad+[[\beta^{(0)},\ldots,\beta^{(j-1)},\delta^{(j+1)},\ldots,\delta^{(j+l)},
\alpha^{(j+l+1)},\ldots,\alpha^{(j+l+k)},\alpha^{(j)}]\sigma,e_n] \cr
&\qquad=[[\beta^{j+1},\delta^l,\alpha^k],e_n]+
q^{2(l+k)}[[\beta^j,\delta^l,\alpha^{k+1}],e_{n+2}].
}}
We use $\rho_{j+k+l+1}(\sigma)=(-q)^{2(k+l)}B^2$ and absorb $(-1)^{l+k}$ by an
isometry of $Q_{j+l+k+1}$. The ordering of loops according to deformation and
the assignement of the components of $(t_0,\ldots,t_{j+k+l})\in Q_{j+l+k+1}$
to the individual loops indicated by superscripts should be clear from the
notation used in \IIIvii. Iterate \IIIvii\ to obtain
\eqn\IIIviii{
[[\delta^j,\alpha^k],e_n]=
\sum_{s=0}^{\min\{j,p-k-1\}}c_s(j,k)[[\beta^{j-s},\alpha^{k+s}],e_{n+2s}]
}
with coefficients
\eqn\IIIix{
c_s(j,k)=\sum_{0\leq i_1\leq\cdots\leq i_s\leq j-s}\prod_{l=1}^s
q^{2(k+j-i_l-1)}=q^{s(j+2k+s-2)}\qbin js
}
using Gau\ss's formula
\eqn\IIIx{
\sum_{0\leq i_1\leq\cdots\leq i_s\leq j-s}q^{-2\sum_{l=1}^s i_l}
=q^{s(s-j)}\qbin js.
}
The $q$-binomial coefficient is defined as
\eqn\IIIxi{
\qbin js :={[j]_q!\over[s]_q![j-s]_q!}.
}
Putting \IIIv,\IIIvi\ , and \IIIviii\ together, it follows that
\eqn\IIIxii{
L_r(T_\alpha)\bigl([[\beta^j,\alpha^k],e_n]\bigr)
=dq^{-{1\over2}n^2}\sum_{s=0}^{\min\{j,p-k-1\}}q^{s(j+2k+s-2)}\qbin js
[[\beta^{j-s},\alpha^{k+s}],e_{n+2s}]
}
with
\eqn\IIIxiii{
d=\sum_{l=0}^{2p-1}q^{{1\over2}l^2}.
}
Thus \IIIxii\ gives the matrix elements of $L_r(T_\alpha)$ in terms of the
basis with elements $[[\beta^j,\alpha^k],e_n]$.

\bigskip
\noindent{\it Action of $T_\beta$}
\bigskip

Let $r=j+k$. The action of $T_\beta$ has the form
\eqn\IIIxiv{
L_r(T_\beta)\bigl([[\beta^j,\alpha^k],e_n]\bigr)
=[[\beta^j,\delta^k],D(T_\beta)e_n],
}
which we again will express in terms of the basis of $A_r(X_r,X_r^0;L_r)$.
Using \IIIvii\ it follows that
\eqn\IIIxv{
[[\beta^j,\delta^k],e_n]=\sum_{s=\max\{0,j+k-p+1\}}^k b_s(j,k)
[[\beta^{j+k-s},\alpha^s],e_{n+2s}]
}
with coefficients
\eqn\IIIxvi{
b_s(j,k)=\sum_{0\leq i_1\leq\cdots\leq i_s\leq k-s}\prod_{l=1}^s
q^{2(k-i_l-1)}=q^{s(k+s-2)}\qbin ks.
}
Note that $b_s(j,k)=c_s(k,0)$. \IIIxiv,\IIIxv\ , and \IIIxvi\ yield
\eqn\IIIxvii{
\eqalign{
&L_r(T_\beta)\bigl([\beta^j,\alpha^k],e_n]\bigr) \cr
&\qquad
=\sum_{l=0}^{2p-1}\sum^k_{s=\max\{0,j+k-p+1\}}q^{-{1\over2}l(l+2)+s(k+s-2)}
\qbin ks [[\beta^{j+k-s},\alpha^s],e_{n+l+2s}],
}}
completing the calculation of the action of $T_\beta$ on $A_r(X_r,X_r^0;L_r)$.

\subsec{Action of ${\cal M}_{1,1}(X)$ on $\uq$}

We have identified $A_r(X_r,X_r^0;L_r)$ as a $\uq$ module as $\uq$ with
the adjoint action. The action of ${\cal M}_{1,1}(X)$ on $A_r(X_r,X_r^0;L_r)$
commutes with the topological action of $\uq$. In this section, we identify
the action of ${\cal M}_{1,1}(X)$ on $\uq$ defined by its action on
$A_r(X_r,X_r^0;L_r)$. By construction it commutes with the adjoint action.

\bigskip
\noindent{\it Action of $T_\alpha$}
\bigskip

Let $\phi_r:A_r(X_r,X_r^0;L_r)\to\uq$ be the restriction of the map defined
in chapter 3. Then
\eqn\IIIxviii{
\eqalign{
&\phi_r\left\{L_r(T_\alpha)\bigl([[\beta^j,\alpha^k],e_n]\bigr)\right\} \cr
&\qquad=dq^{-{1\over2}n^2}\sum_{s=0}^{\min\{j,p-k+1\}}q^{s(j+2k+s-2)}\qbin js
\cr
&\qquad\quad\times N(j-s,k+s,n+2s)F^{k+s}T_{n+2s-1}E^{p-j+s-1},
}}
using \IIIxii. We define $U(T_\alpha):\uq\to\uq$ by
\eqn\IIIxix{
U(T_\alpha)\circ\phi_r=\phi_r\circ L_r(T_\alpha).
}
Thus, what is left to compute $U(T_\alpha)$, is the ratio of normalization
constants
\eqn\IIIxx{
{N(j-s,k+s,n+2s)\over N(j,k,n)}=
(-1)^sq^{-s(j+2k+{3\over2}s+n-{3\over2})}[1]_q^s{[j-s]_q!\over[j]_q!}.
}
We conclude that
\eqn\IIIxxi{
\eqalign{
&U(T_\alpha)(F^kT_{n-1}E^{p-j-1}) \cr
&\qquad=dq^{-{1\over2}n^2}\sum_{s=0}^{\min\{j,p-1-k\}}(-1)^s
q^{-{1\over2}s(s+1)-sn}{[1]_q^s\over[s]_q!}F^{k+s}T_{n+2s-1}E^{p-j+s-1},
}}
giving the action of $U(T_\alpha)$ on $\uq$.

\bigskip
\noindent{\it Action of $T_\beta$}
\bigskip

Let $r=j+k$. Using \IIIxvii, we obtain
\eqn\IIIxxii{
\eqalign{
&\phi_r\left\{L_r(T_\beta)\bigl([[\beta^j,\alpha^k],e_n]\bigr)\right\} \cr
&\qquad=\sum_{l=0}^{2p-1}\sum^k_{s=\max\{0,j+k-p+1\}}
q^{-{1\over2}l(l+2)+s(k+s-2)}\qbin ks \cr
&\qquad\quad\times N(j+k-s,s,n+l+2s)F^sT_{n+l+2s-1}E^{p-j-k+s-1}.
}}
Insert
\eqn\IIIxxiii{
\eqalign{
&{N(j+k-s,s,n+l+2s)\over N(j,k,n)} \cr
&\qquad={(-1)^{k-s}\over[1]_q^{k-s}}
q^{{1\over2}(k-s)(k-s+1)+(k-s)(j+n+2)+(2s+l)(2-s)}
{[j+k-s]_q!\over[j]_q!}
}}
to conclude that
\eqn\IIIxxiv{
\eqalign{
&U(T_\beta)(F^kT_{n-1}E^{p-j-1}) \cr
&\qquad=\sum_{l=0}^{2p-1}
\sum^k_{s=\max\{0,j+k-p+1\}}{(-1)^{k-s}\over[1]_q^{k-s}}
q^{{1\over2}(k-s)(k-s+1)+(k-s)(j+n+s+2)+{1\over2}(l+2s)(1-l)} \cr
&\qquad\quad\times[k-s]_q!\qbin{j+k-s}j\qbin ks
F^sT_{n+l+2s-1}E^{p-j-k+s-1}
}}
completing the calculation of $U(T_\beta)$.

\subsec{Action of $S_{\alpha\beta}$}

We define $S_{\alpha\beta}:=T_\beta(T_\alpha)^{-1}T_\beta$. The action of
$S_{\alpha\beta}$ on $\pi_1(X,p_0)$ is given by $S_{\alpha\beta}(\alpha)=\beta$
and $S_{\alpha\beta}(\beta)=\alpha'$. Recall that $\alpha'=\beta^{-1}\circ
\alpha^{-1}\circ\beta$. That is $S_{\alpha\beta}$ maps $\alpha$ to $\beta$
and $\beta$ to $\alpha^{-1}$ conjugated by $\beta^{-1}$. This transformation
is known as the $S$-transformation. We put
\eqn\IIIxxv{
\eqalign{
&D(S_{\alpha\beta}):=D(T_\beta)D(T_\alpha)^{-1}D(T_\beta), \cr
&D(S_{\alpha\beta})_{m,n}={2pq\over d^2}q^{(m+1)(n-1)}.
}}
$D(S_{\alpha\beta})$ performs a discrete Fourier transformation on $V$.
$\rho_r:\pi_1(X_r,*_r)\to \GL(V)$ is compatible with $S_{\alpha\beta}$,
the equivalence being given by $D(S_{\alpha\beta})$. We thus have an action
$L_r(S_{\alpha\beta})$ on $A_r(X_r,X_r^0;L_r)$. It has the form
\eqn\IIIxxvi{
L_r(S_{\alpha\beta})\bigl([[\beta^j,\alpha^k],e_n]\bigr)
=[[\alpha'^j,\beta^k],D(S_{\alpha\beta})e_n],
}
$r=j+k$. As before, we deduce from \IIIxxvi\ an action $U(S_{\alpha\beta})$
on $\uq$. However, we do not expand \IIIxxvi\ in terms of the basis of
$A_r(X_r,X_r^0;L_r)$ as we did in the case of $L_r(T_\alpha)$ and
$L_r(T_\beta)$, although this could be done. Instead we compute directly the
image of \IIIxxvi\ under $\phi_r$. Using
\eqn\IIIxxvii{
[[\alpha'^j,\beta^k],e_n]=q^{j(j+2k+1)}(F_R)^j[[\beta^k],e_n]
}
it follows that
\eqn\IIIxxviii{
\eqalign{
&\phi_r\bigl([[\alpha'^j,\beta^k],e_n]\bigr) \cr
&\qquad=(-1)^k[1]_q^{j-k}{[k]_q!\over[j]_q!}
q^{{1\over2}j(j+1)+{1\over2}k(k+1)+2k(j+1)+n(j+k)} \cr
&\qquad\quad\times N(j,k,n)T_{n-1}E^{p-k-1}F^j.
}}
Note that by reexpressing \IIIxxviii\ in terms of the basis
$F^jT_{k-1}E^{p-l-1}$
and applying $(\phi_r)^{-1}$, the expansion of \IIIxxvi\ in terms of the basis
of $A_r(X_r,X_r^0;L_r)$ could be obtained. We conclude that
\eqn\IIIxxix{
\eqalign{
&\phi_r\left\{L_r(S_{\alpha\beta})\bigl([[\beta^j,\alpha^k],e_n]\bigr)\right\}
\cr
&\qquad={2pq\over d^2}(-1)^k[1]_q^{j-k}{[k]_q!\over[j]_q!}
q^{{1\over2}j(j+3)+{1\over2}k(k+3)+k(n+1+2j)} \cr
&\qquad\quad\times N(j,k,n)K^{2(j+n+1)}E^{p-k-1}F^j,
}}
using
\eqn\IIIxxx{
\sum_{l=0}^{2p-1}q^{(n+l-1)(j+n-1)}T_{n+l-1}=K^{2(j+n+1)}.
}
The final result is
\eqn\IIIxxxi{
\eqalign{
&U(S_{\alpha\beta})(F^kT_{n-1}E^{p-j-1}) \cr
&\qquad={2pq\over d^2}(-1)^k[1]_q^{j-k}{[k]_q!\over[j]_q!}
q^{{1\over2}j(j+3)+{1\over2}k(k+3)+k(n+1+2j)}
K^{2(j+n+1)}E^{p-k-1}F^j.
}}
$U(S_{\alpha\beta})$ is the algebraic version of the $S$-transformation. It is
a mapping of $\uq$ to itself, one-to-one and onto, which commutes with the
adjoint action.

\newsec{Identification of the $S$- and the $T$-transformation}

We identify the operations $U(T_{\alpha})$ and
$U(S_{\alpha\beta})$ in terms of the quasitriangular structure of
$\uq$. Let us first adjust the normalization of
$D(T_{\alpha})$ and $D(S_{\alpha\beta})$ as follows:
\eqn\IVi{
\eqalign{
D(T_{\alpha})&\rightarrow {N_\alpha\over d} D(T_{\alpha}), \cr
D(S_{\alpha\beta})&\rightarrow {{d^{2}N_{\alpha\beta}}\over {2pq}}
D(S_{\alpha\beta}).
}}
With this change of normalization, $U(T_{\alpha})$ and
$U(S_{\alpha\beta})$ act on $\uq$ by
\eqn\IVii{
\eqalign{
&U(T_\alpha)(F^kT_{n-1}E^{p-j-1}) \cr
&\qquad=N_{\alpha}q^{-{1\over2}n^2}\sum_{s=0}^{\min\{j,p-1-k\}}(-1)^s
q^{-{1\over2}s(s+1)-sn}{[1]_q^s\over[s]_q!}F^{k+s}T_{n+2s-1}E^{p-j+s-1},
}}
and
\eqn\IViii{
\eqalign{
&U(S_{\alpha\beta})(F^kT_{n-1}E^{p-j-1}) \cr
&\qquad=N_{\alpha\beta}(-1)^k[1]_q^{j-k}{[k]_q!\over[j]_q!}
q^{{1\over2}j(j+3)+{1\over2}k(k+3)+k(n+1+2j)}
K^{2(j+n+1)}E^{p-k-1}F^j.
}}

\subsec{Universal elements of $\uqk$}

Let us consider the $K$ generated version of $\uq$.
It is known to be a ribbon Hopf algebra. The universal $R$-matrix is
\eqn\IViv{
\eqalign{
R={1\over 4p}\left(\sum_{n=0}^{p-1}(-1)^{n}
{{[1]_{q}^{n}}\over{[n]_{q}!}}
q^{- {1\over2} n(n-1)} E^{n}\otimes F^{n}\right)
\left(\sum_{m,n=0}^{4p-1}q^{{1\over2} nm}K^{n}\otimes K^{m}\right).
}}
The associated central element $V$ is
\eqn\IVivbis{
\eqalign{
V&=\sum_{n=0}^{p-1}\sum_{m=0}^{4p-1}
(-1)^{n}{[1]_{q}^{n}\over[n]_{q}!}
q^{{1\over2} n(n+1)+n(m+1)+{1\over2}m(m+2)}
F^{n}H_{m+2n}E^{n}, \cr
H_n &={1\over 4p}\sum_{m=0}^{4p-1}q^{-{1\over2}nm}K^m.
}}

\subsec{Identification of $U(T_{\alpha})$}

Let $N_\alpha=q^{1\over2}$. Then
\eqn\IVv{
\def\mapright#1{\smash{\mathop{\longrightarrow}\limits^{#1}}}
\def\mapdown#1{\Big\downarrow\rlap{$\vcenter{\hbox{$\scriptstyle#1$}}$}}
\matrix{
\uq &\mapright{U(T_\alpha)} &\uq \cr
\mapdown{} && \mapdown{} \cr
\uqk &\mapright{\lambda_{V^{-1}}} &\uqk
}}
commutes. That is, $U(T_\alpha)$ is identified with the multiplication
by the inverse of the central element $V$ of \IViv. This is shown by
\eqn\IVvi{
\eqalign{
&V^{-1}F^kT_{n-1}E^{p-j-1} \cr
 &\qquad=\sum_{s=0}^{p-1}\sum_{l=0}^{4p-1}(-1)^s{[1]_q^s\over[s]_q!}
q^{-{1\over2}s(s+1)-sl-{1\over2}l^2+{1\over2}}
F^{k+s}H_{l+2s-1}T_{n+2s-1}E^{p-1-j+s}
}}
with
\eqn\IVvii{
\eqalign{
H_{l+2s-1}T_{n+2s-1}
&=H_{l+2s-1}(H_{n+2s-1}+H_{n+2p+2s-1}) \cr
&=\delta_{l,n}H_{n+2s-1}+\delta_{l,n+2p}H_{n+2p+2s-1},
}}
comparing the result with \IVii.

\subsec{Trace on $\uqk$}

Let $\tau:\uqk\to\C$ be the linear map such that
\eqn\IVviii{
\eqalign{
\tau(E^{p-1}F^{p-1}H_{n-1})&:=1,\qquad 1\leq n\leq 4p, \cr
\tau(E^jF^kH_{n-1})&:=0,\qquad\hbox{\rm else}.
}}
$\tau$ is a trace on $\uqk$: For $X=E,F,K^{\pm 1}$, and $Y\in\uqk$,
$\tau(XY)=\tau(YX)$.

\subsec{$S$-transformation}

Let $R=\sum_{i}\alpha_i\otimes\beta_i$ the universal $R$-matrix \IViv.
We define a linear map $S:\uqk\to\uqk$ by
\eqn\IVix{
S(X):=\sum_{j,l}\beta_j\eta(\alpha_l)\tau\bigl(\eta(\beta_l)K^{-2}
\alpha_jX\bigr)
}
with $\tau$ the trace of \IVviii. A short computation reveals that
\eqn\IVx{
\eqalign{
&S(T_{n-1}E^{p-1-r}F^{p-1-s}) \cr
&\qquad={2(-1)^{r+s}\over[r]_q![s]_q!}
q^{-{1\over2}r(r+1)-{1\over2}s(s+1)+(s-1)(2r+n-1)}
F^rK^{2(1-n-r)}E^s.
}}
We thus obtain a map $S:\uq\to\uq$ by restriction.

\subsec{Identification of $S_{\alpha\beta}$}

Put the normalization constant in \IVi\ to be
\eqn\IVxi{
N_{\alpha\beta}:={1\over 2p}{[p-1]_q!\over[1]_q^{p-1}}
q^{-{1\over2}(p-1)(p+2)}.
}
The transformation \IViii\ is identified as
\eqn\IVxii{
U(S_{\alpha\beta})^{-1}X=S(X)
}
with $S$ the transformation \IVx. To verify \IVxii, we compute the
inverse transformation of \IViii. it is seen to be
\eqn\IVxiii{
\eqalign{
&U(S_{\alpha\beta})^{-1}(T_{n-1}E^{p-1-k}F^j) \cr
&\qquad={1\over pN_{\alpha\beta}}(-1)^k[1]_q^{k-j}{[j]_q!\over[k]_q!}
q^{-{1\over2}j(j+3)-{1\over2}k(k+3)-jk-(n+k-1)(j+2)}
F^kK^{-2(n+k-1)}E^{p-j-1}.
}}
Setting $k=r$ and $j=p-1-s$ and comparing \IVx\ with \IVxiii, the
result \IVxii\ follows.

\listrefs
\end